\documentclass{article}
\usepackage{amsmath}
\usepackage{epsfig}
\usepackage{dcolumn}

\psfull

\begin{document}

\title{Analysis of the optimality principles responsible for vascular network
architectonics}
\author{I.A.Lubashevsky, V.V.Gafiychuk \\
Institute of General Physics Academy \\
of Science of Russian Federation; \\
Moscow State University, Moscow\\
Institute of Applied Problems of Mechanics and Mathematics \\
National Academy of Sciences of Ukraine, Lviv}
\maketitle

\begin{abstract}
The equivalence of two optimality principles leading to Murray's law has
been discussed. The first approach is based on minimization of biological
work needed for maintaining the blood flow through the vessels at required
level. The second one is the principle of minimal drag and lumen volume.
Characteristic features of these principles are considered.

An alternative approach leading to Murray's law has been proposed. For that
we model the microcirculatory bed in terms of delivering vascular network
with symmetrical bifurcation nodes, embedded uniformly into the cellular
tissue. It was shown that Murray's law can be regarded as a direct
consequence of the organism capacity for controlling the blood flow
redistribution over the microcirculatory beds.
\end{abstract}

\section{Introduction}

\noindent\ The great amount of natural systems have highly branching
networks. As a evident example of such systems we may regard living tissue
where blood supplies the cellular tissue with oxygen, nutritious products,
etc. through branching vascular network and at the same time withdraws
products resulting from living activities of the cellular tissue. A similar
situation takes place in respiratory systems where oxygen reaches small
vessels (capillaries) going through the hierarchical system of bronchial
tubes. There is a question of what physical principles govern network
organization living systems under consideration. In this paper we focus our
attention on the analyzes of different known optimal principles of network
formation for microcirculatory bed and developing new approach to this
problem.\bigskip

\section{Analysis of the optimality principles}

A microcirculatory bed can be reasonably regarded as a space-filling fractal
being a natural structure for ensuring that all cells are serviced by
capillaries \cite{WBE97}. The vessel network must branch so that every small
group of cells, referred below to as ``elementary tissue domain'', is
supplied by at least one capillary. Since a typical length of capillaries is
about 0.3 to 0.7 mm a vessel network generated by an artery of length of
order of 1 to 5 cm should contain a sufficiently large number of hierarchy
levels. At zeroth level we meet the host artery and the host vein, the
mother and daughter vessels belong to $n$-th and $(n+1)$-th levels,
respectively, and the last level $N$ comprises capillaries. So at each level
$n$ of the vascular network the tissue domain supplied by a given
microcirculatory bed as a whole can be approximated by the union of the
tissue subdomains whose mean size is about the typical length $l_{n}$ of the
$n$-th level vessels. Thus, the individual volume of these subdomains is
estimated as $V_{n}\sim l_{n}^{3}$ and their total number (as well as the
total number of $n$-th level vessels) is about $M_{n}\sim V_{0}/l_{n}^{3}$,
where $V_{0}$ is the total volume of the microcirculatory bed. The higher is
the level, the more accurate become the independence of such estimates from
the particular details of vessel arrangements. For internal organs they
approximately hold also for large vessels of regional circulation. To
justify the latter statement we present Table~\ref{T1.3} relating the vessel
lengths and radii to the radii of the corresponding tissue cylinders, i.e.
the cylindrical neighborhood falling on one vessel of a fixed level.

\begin{table}
\caption[Typical parameters of the vessel arrangement]
{Typical parameters of the vessel arrangement of a 13-kg dog,
after~\protect\cite{WJL84}.\label{T1.3}}
\begin{center}
\begin{tabular}{|l|D{.}{.}{3}D{.}{.}{4}D{.}{.}{5}D{.}{.}{7}|}\hline
\vphantom{\Huge V}Vessel type
&
\multicolumn{1}{r}{\hspace{-8pt}\begin{tabular}[c]{c}
Diameter\\$2a$, $\mu$m\end{tabular}}
&
\multicolumn{1}{r}{\hspace{-10pt}\begin{tabular}[c]{c}
Length\\$l$, cm\end{tabular}}
&
\multicolumn{1}{r}{\hspace{-10pt}\begin{tabular}[c]{l}
Tissue/Vessel\\radii,\quad $d/a$\end{tabular}}
&
\multicolumn{1}{r|}{\hspace{-10pt}\begin{tabular}[c]{l}
Arrangement\\anisotropy$^{*}$\end{tabular}}
\\\hline\hline
\vphantom{\LARGE P}Primary arteries& 300 & 1.0 & 30 & 0.86 \\
Small arteries & 100 & 0.5 & 20 & 0.50 \\
Terminal vessels
& 50 & 0.2 & 10 & 0.37 \\
Arterioles & 20 & 0.1 & 7 & 0.25 \\[3pt] \hline
\multicolumn{5}{l}{\vphantom{\LARGE M}${}^{\ast}$measured
as $(\pi d^{2}l)^{1/3}/l$}
\end{tabular}
\end{center}
\end{table}

This condition that the vascular network be volume-preserving from one
generation to the next gives us immediately the local relation between the
characteristic lengths of the vessels: $l_{n}^{3}\approx gl_{n+1}^{3}$ (here
$g=2$ is the order of the vessel branching node). Whence it follows that $%
l_{n}\sim l_{0}g^{-n/3}$, where $l_{0}$ is the characteristic size of the
microcirculatory bed region or, what is practically the same, the length of
the host artery.

\begin{figure}[t]
\begin{center}
\psfig{file=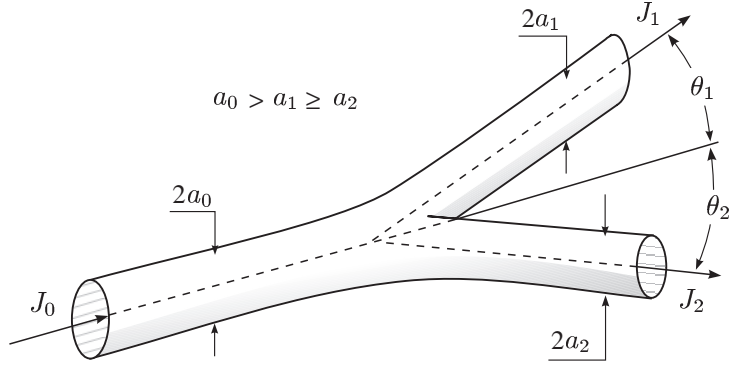}
\end{center}
\caption{Characteristics of the vessel branching}
\label{F1.5}
\end{figure}

The following analysis, however, will require a more detailed information
about the vascular network architectonics. Namely, we need to know how the
vessel radii change at the nodes and the relative arrangement of mother and
daughter vessels. Actually here we meet the problem as to what fundamental
regularities govern the vessel branching. These regularities manifest
themselves in the relation between the radii $a_{0}$, $a_{1}$, $a_{2}$ of
mother and daughter arteries, respectively, and the angles $\theta _{1}$, $%
\theta _{2}$, $\theta _{12}=\theta _{1}+\theta _{2}$ which the daughter
branches make with the direction of the mother artery and with each other
(Fig.~\ref{F1.5}).

\begin{table}
\caption[Integral characteristics of the vessel cross-section]
{Integral characteristics of the vessel cross-section at different branching
levels, after~\protect\cite{LaB90}.\label{T1.4}}
\begin{center}
\begin{tabular}{|l|D{.}{.}{3}D{.}{.}{2}D{.}{.}{5}l|}
\hline
Vessel type\rule[-10pt]{0mm}{27.5pt}
&
\multicolumn{1}{r}{\hspace{-4pt}
\begin{tabular}[c]{c}
Mean radius\\ $a$, $\mu$m\end{tabular}}
&
\multicolumn{1}{r}{\hspace{-4pt}
\begin{tabular}[c]{c}
$\sum a^{2}$\\cm$^{2}$\end{tabular}}
&
\multicolumn{1}{r}{\hspace{-4pt}
\begin{tabular}[c]{c}
$\sum a^{3}\cdot 10$\\cm$^{3}$\end{tabular}}
&
\multicolumn{1}{c|}{%\hspace{-4pt}
\begin{tabular}[c]{c}
$\sum a^{4}$\\cm$^{4}$\end{tabular}}
\\\hline\hline
&\multicolumn{4}{c|}{\vphantom{\LARGE H}\textit{\small Homo sapiens}} \\[3pt]
Aorta & 12500  & 1.56 & 1.95 & 2.44 \\
Arteries & 2000 & 6.36 & 1.27 & 0.25 \\
Arterioles & 30 & 127.4 & 0.382 & 1.15${}\times 10^{-3}$ \\
Capillaries & 6 & 1432 & 0.860 & 5.16${}\times 10^{-4}$ \\
Venules & 20 & 1273 & 2.55 & 5.09${}\times 10^{-3}$ \\
Veins & 2500 & 12.9 & 3.18 & 0.80 \\
Vena cava & 15000 & 2.25 & 3.38 & 5.06 \\
&\multicolumn{4}{c|}{\vphantom{\LARGE H}\textit{\small Canis familiaris}} \\[3pt]
Aorta & 5000 &  & 1.25 &  \\
Large arteries & 1500 &  & 1.35 &  \\
Main arterial  & 500 &  & 0.75 &  \\[-3pt]
branches &\multicolumn{4}{c|}{}\\
Terminal branches & 300 &  & 0.486 &  \\
Arteriolas & 10 &  & 0.400 &  \\
Capillaries & 4 &  & 0.768 &  \\
Venules & 15 &  & 2.70 &  \\
Terminal veins & 750 &  & 7.59 &  \\
Main venous & 1200 &  & 10.37 &  \\[-3pt]
branches &\multicolumn{4}{c|}{}\\
Large veins & 3000 &  & 10.80 &  \\
Vena cava & 6250 &  & 1.53 &  \\ \hline
\end{tabular}
\end{center}
\end{table}

A detailed theoretical attempt of understanding this regularity was made
first by Cecil~D.~Murray in 1926 \cite{M26a,M26b}. He proposed a model
relating the artery radii at branching nodes (Fig.~\ref{F1.5}) by the
expression 
\begin{equation}
a_{0}^{x}=a_{1}^{x}+a_{2}^{x}\quad \text{with}\quad x=3  \label{e1.2:1}
\end{equation}
thereafter referred to as Murray's law ($x$ is also called the bifurcation
exponent). Then Murray's approach was under development in a large number of
works, see, for example, \cite{C54,C55,R67,Sho82} and a series of works by
Zamir \textit{et all.} \cite{z76a,z76b,z88a,z88b} and by Woldenberg \textit{%
et all.} \cite{WH86} (for a historical review see also \cite{S81,z88a,LaB90}%
). The idea of Murray's model is reduced to the assumption that
physiological vascular network, subject through evolution to natural
selection, must have achieved an optimum arrangement corresponding to the
least possible biological work needed for maintaining the blood flow through
it at required level. This biological work $\mathcal{P}$ involves two terms:
(\textit{i}) the cost of overcoming viscous drag during blood motion through
the vessels obeying Poiseuille's law, and (\textit{ii}) the energy
metabolically required to maintain the volume of blood and the vessel
tissue. Dealing with an individual artery of length $l$ and radius $a$ with
a blood flow rate $J$ in it we get: 
\begin{equation}
\mathcal{P}=\frac{8\eta lJ^{2}}{\pi a^{4}}+m\pi a^{2}l\,,  \label{e1.2:2}
\end{equation}
where $\eta $ is the blood viscosity and $m$ is a metabolic coefficient.
Minimizing function~(\ref{e1.2:1}) with respect to $a$ we find the relation
between the blood flow rate $J$ and the artery lumen radius $a$
corresponding to the given optimality principles: 
\begin{equation}
J=ka^{3},  \label{e1.2:3}
\end{equation}
where the coefficient $k=\sqrt{m\pi ^{2}/(16\eta )}$ is a constant for the
tissue under consideration. Due to the blood conservation at branching nodes
we can write $J_{0}=J_{1}+J_{2}$ (Fig.~\ref{F1.5}) whence Murray's law~(\ref
{e1.2:1}) immediately follows.

However, care must be taken in comparing measurements with prediction,
particularly if averages over many successive levels are used. Already Mall
himself noted that his data were approximate \cite{M88}. In particular, for
large arteries of systemic circulation where blood flow can be turbulent the
bifurcation exponent $x$ should be equal to $7/3\approx 2.33$ as it follows
from this optimality principle of minimum pumping power and lumen volume.

\begin{figure}[tbp]
\begin{center}
\psfig{file=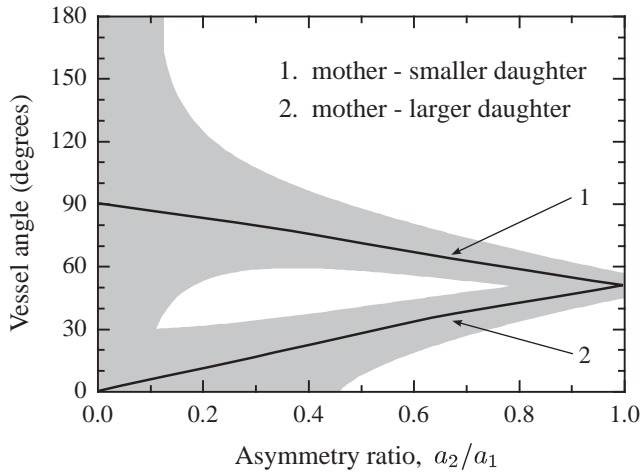}
\end{center}
\caption{The angles that the daughter arteries make with the mother artery
when the vessel branching is governed by the minimum drag \& suface
principle (solid lines) and under 10\% perturbations (darkened region).}
\label{F1.6}
\end{figure}

There is also another optimality principle leading to Murray's law, the
principle of minimum drag and lumen surface \cite{z76a,z76b}. The drag
against the blood motion through vessels is caused by the blood viscosity
and can be described in terms of the shear stress on the walls of vessels, $%
2\pi al\tau _{t}$, where 
\begin{equation}
\tau _{t}=\eta \nabla _{n}v=\frac{a}{2l}\delta P=\frac{4\eta }{\pi }\,\frac{J%
}{a^{3}}  \label{e1.2:4}
\end{equation}
for the laminar flow and $\delta P$ is the pressure drop along a vessel of
length $l$ and radius $a$. Then the given optimality principle is reduced to
the minimum condition for the function 
\begin{equation}
\mathcal{P}^{\prime }=\frac{8\eta lJ}{a^{2}}+m^{\prime }2\pi al\,,
\label{e1.2:5}
\end{equation}
where $m^{\prime }$ is a certain weighting coefficient. Minimizing (\ref
{e1.2:5}) with respect to $a$ we get a relationship between $J$ and $a$ of
the same form as (\ref{e1.2:3}), leading to Murray's law again. There were a
number works (see, e.g., \cite{z88a,WH86,vBS92,GE90}) aimed at finding out
what the specific optimality principle governs the artery branching by
studying the angles of daughter vessels, $\theta _{1}$, $\theta _{2}$, $%
\theta _{12}$, in relation to the asymmetry of the branching node, $%
a_{2}/a_{1}$ (Fig.~\ref{F1.5}). However, on one hand, all the optimality
principles give numerically close relationships between the vessel angles
and radii for the bifurcation exponent $x\approx 3$ \cite{WH86}. On the
other hand, it turned out that experimentally determined branching angles
generally exhibit considerable scatter around the theoretical optimum. The
matter is that small variations of the total ``cost'' of artery bifurcation
about several percents causes the actual vessel angles to deviate
significantly from the predicted optimum. This feature is illustrated in
Fig.~\ref{F1.6} showing the variations in the vessel angles governed by the
minimality of functional~(\ref{e1.2:5}) with imposed 10\% perturbations.
Namely, varying the coordinates of the branching node (Fig.~\ref{F1.5}) we
get that the minimum of the function $\mathcal{P}_{0}^{\prime }+\mathcal{P}%
_{1}^{\prime }+\mathcal{P}_{2}^{\prime }$ is attained when 
\begin{eqnarray}  \label{e1.2:7}
a_{1}\cos \theta _{1}+a_{2}\cos \theta _{2} &=&(1+\epsilon )a_{0}\,,
\label{e1.2:6} \\
a_{1}\sin \theta _{1}-a_{2}\sin \theta _{2} &=&\epsilon ^{\prime }a_{0}\,,
\label{e1.2:7}
\end{eqnarray}
where the additional terms $\epsilon a_{0}$ and $\epsilon ^{\prime }a_{0}$
with $\left| \epsilon \right| ,\left| \epsilon ^{\prime }\right| <0.1$
describes possible deviations from the optimality condition. The resulting
values of $\theta _{1}$ and $\theta _{2}$ are depicted in Fig.~\ref{F1.6}.
It should be noted that expressions~(\ref{e1.2:6}), (\ref{e1.2:7})
correspond actually to the mechanical equilibrium of the node under the
action of vessel walls strained by the blood motion and the additional terms
describe a possible effect of the cellular tissue.

Nevertheless, the optimality principles based on functional~(\ref{e1.2:2})
seems to govern the artery bifurcations \cite{GE90}. Besides, this
principles gives also adequate estimates of the integral characteristics of
microcirculatory beds \cite{KB91,KB94}.

The bifurcation exponent $x$, on the contrary, is well approximated by the
Murray value, $x\approx 3$, at least starting from arteries of intermediate
size \cite{z83}. This value meets also the space-filling requirement for the
vascular network fractal in geometry to fill precisely the space of a fixed
relative volume at each hierarchy level \cite{Mf82}. Indeed, assuming the
volume of the tissue cylinders matching an artery of length $l$ and lumen
radius $a$ to be about $l^{3}$ we get that the corresponding relative volume
of blood is $(a/l)^{2}$. So it is fixed if $a=\mathrm{constant}\cdot l$ and,
thus, $a_{0}^{3}=a_{1}^{3}+a_{2}^{3}$ provided the tissue cylinder matching
the mother artery is composed of the tissue cylinders of the daughter
arteries.

\begin{figure}[tbp]
\begin{center}
\psfig{file=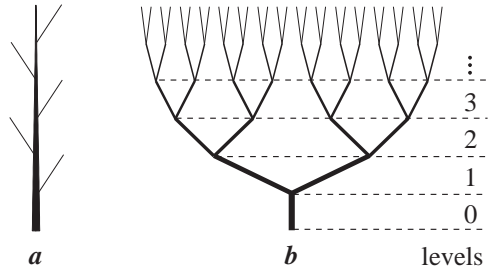}
\end{center}
\caption{Schematic illustration of the ``conveying'' ({\protect\boldmath$a$}%
) and ``deliverying'' ({\protect\boldmath$b$}) types of artery pattern.}
\label{F1.7}
\end{figure}

In order to specify the microcirculatory bed structure we need also to
classify vessels according to the symmetry of their branching (Fig.~\ref
{F1.7}). The matter is that \cite{z88b} arteries with predominantly
asymmetric bifurcations give off comparatively little flow into its side
branches along its way and, therefore, able to carry the mainstream flow
across larger distances. Conversely, a more symmetric bifurcation pattern
splits flow into numerous small branches, thereby delivering blood to its
surrounding tissue. Such arteries have been attributed a ``conveying'' and
``delivering'' types of function, respectively. Since blood must be conveyed
towards the sites at which to be delivered, both types of vessels occur in
real arterial trees. Moreover, a larger conveying vessel may switch into a
bunch of small delivering branches.

\begin{figure}[tbp]
\begin{center}
\psfig{file=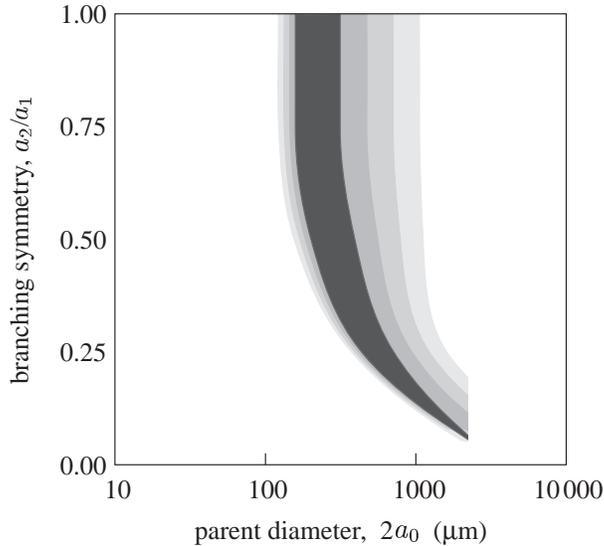}
\end{center}
\caption{Bifurcation symmetry \textit{vs.} vessel diameter for the porcine
coronary arterial tree (based on the data of \protect\cite{vBS92}.) The
darkness intensity indicates the density of the experimental points.}
\label{F1.8}
\end{figure}

Obviously, real arterial trees should contain a great variety of
intermediate stages in between these extremes and as our field of view moves
from the large systemic arteries to small arteries of regional circulation
the vessel bifurcation should become more and more symmetrical. This has
been also justified by numerically modelling the structure of arterial trees
governed by the minimality condition of blood volume \cite
{S93,SB93,SNNREB94,SNNEM96}. According to the experimental data (see, e.g.,
the work~\cite{vBS92} and Fig.~\ref{F1.8} based on it) even sufficiently
large regional arteries of diameter and length about 300~$\mu $m and 1~cm,
respectively, (Table~\ref{T1.3}) branch symmetrically, at least at first
approximation.. Therefore microcirculatory beds as they have been specified
above can be regarded as a vessel network with approximately symmetrical
bifurcations.

In other words, we may think of the systemic arteries as vessels of the
conveying type where the mean blood pressure is practically constant.
Conversely, the arteries of microcirculatory beds should belong to the
delivering type and mainly determine the total resistance of the vascular
network to blood flow, with the blood pressure drop being uniformly
distributed over many arteries of different length.

\section{Physiological mechanisms governing the vessel arrangements}

The universality of Murray's law for distributed transport systems in many
different live organisms raises questions as to: What cues are available to
organisms to use in generating such systems? What physiological mechanisms
enable them to adapt to altering conditions? Do in fact live organisms
follow certain global optimality principles?

For Murray's law~(\ref{e1.2:3}) the shear stress $\tau _{t}$ is constant
(see formula~(\ref{e1.2:4})) throughout a given artery system. Rodbard \cite
{R75} proposed that the shear stress detected by the vessel endothelium
leads to the vessel growth or contraction, and Zamir \cite{z77} suggested
that this leads to the development of Murray's system as vessels maintain a
constant value of shear stress. Concerning the particular mechanism by which
organisms can implement the shear stress sensitivity we can say the
following. Now it is established that the adaptation of conduit arteries as
well as resistance arteries to acute changes in flow is mediated by the
potent endogenous nitrovasodilator endothelium-derived relaxing factor,
whose release from endothelial cells is enhanced by flow through the
physical stimulation of shear stress (see, e.g., \cite{GE90} and references
therein). The adaptation of arterial diameters to long-term changes in the
flow rate also occurs through a mechanism which appears to involve the
sensitivity to shear stress and the participation of endothelial cells, but
remains not to be understood well \cite{GE90}.

It should be noted that the shear stress equality through a vascular network
does not lead directly to a certain optimality principle. Different
principles, for instance, (\ref{e1.2:2}) and (\ref{e1.2:5}), can give the
same condition imposed on the shear stress. Moreover, it is quite possible
that the case of this equality is of another nature. In particular, for
large conduit arteries in the human pulmonary tree the bifurcation exponent $%
x$ is reported to be in the range 1--2, whereas Murray's law holds well
beginning from intermediate conveying arteries \cite{WH83,Li96}. The matter
is that in large systemic arteries the blood pressure exhibits substantial
oscillations because of the heart beating, giving rise to damped waves
travelling through the systemic arteries. The value of the bifurcation
exponent $x=2$ matching the area-preserving law at the branching nodes
ensures that the energy-carrying waves are not reflected back up the vessels
at the nodes. However, this requirement is also can be derived from a
certain optimality principle \cite{WBE97}.

Summarizing the aforesaid we will model the microcirculatory bed in terms of
a delivering vascular network with symmetrical bifurcation nodes embedded
uniformly into the cellular tissue. Besides, the Murray's law will be
assumed to hold. The latter is also essential from the standpoint of the
tissue self-regulation, which will be discussed in detail in the next
section. Here, nevertheless, we make several remarks concerning the given
aspect too, because it could be treated as an alternative origin of Murray's
law~(\ref{e1.2:3}). Let us consider a symmetrical dichotomous vessel tree
shown, e.g., in Fig.~\ref{F1.7}{\boldmath$b$}. In order to govern blood flow
redistribution over the microcirculatory bed finely enough so to supply
with, for example, increased amount of blood only those regions where it is
necessary and not to disturb other regions the blood pressure should be
uniformly distributed over the microcirculatory bed, at least, approximately.

The blood pressure drop $\delta P_{n}$ along an artery of level $n$ ($%
n=0,1,2,\ldots $, Fig.~\ref{F1.7}) for laminar blood flow is
\begin{equation}
\delta P_{n}=\frac{8\eta l_{n}J_{n}}{\pi a_{n}^{4}}\,.  \label{e1.2:8}
\end{equation}
For the space-filling vascular network this artery supplies with blood a
tissue region of volume about $l_{n}^{3}$ and, so, under normal conditions
the blood flow rate $J_{n}$ in it should be equal to $J_{n}\approx
jl_{n}^{3} $, where $j$ is the blood perfusion rate (the volume of blood
flowing through a tissue domain of unit volume per unit time) assumed to be
the same at all the points of the given microcirculatory bed. Then formula~(%
\ref{e1.2:8}) gives us the estimate 
\begin{equation*}
\delta P_{n}=\frac{8\eta j}{\pi }\left( \frac{l_{n}}{a_{n}}\right) ^{4}
\end{equation*}
whence it follows that $\delta P_{n}$ will be approximately the same for all
the levels, i.e. the blood pressure will be uniformly distributed over the
arterial bed if the ratio $l_{n}/a_{n}$ takes a certain fixed value, $%
l_{n}\approx \mathrm{constant}\cdot a_{n}$ and, thus, $J_{n}\approx \mathrm{%
constant}^{\prime }\cdot a_{n}^{3}$. Due to the blood conservation at
branching nodes we can write 
\begin{equation}
J_{0}=J_{1}+J_{2}  \label{10}
\end{equation}
(see Fig.~\ref{F1.5}). The later gives us immediately Murray's law~(\ref
{e1.2:1}). In other words, Murray's law can be also regarded as a direct
consequence of the organism capacity for controlling finely the blood flow
redistribution over the microcirculatory beds.

It should be noted that in the previous papers \cite{gaf1,gaf2,gaf3}\ we
considered in detail the mathematical model for the vascular network
response to variations in the tissue temperature on the given network
architectonics. We have found that the distribution of the blood temperature
over the venous bed aggregating the information of the cellular tissue state
allows the living tissue to function properly. We showed that this property
is one of the general basic characteristics of various natural hierarchical
systems. These systems differ from each other by the specific realization of
such a synergetic mechanism only.

\end{document}